\documentclass[conference, letterpaper]{IEEEtran}
\pdfoutput=1
\usepackage{blindtext, graphicx, caption, varwidth, cite}
\usepackage[ left = .62in, right = .62in, top = .75in, bottom = 1 in]{geometry}
\usepackage{subfigure}
\usepackage{float}
\usepackage{fancyvrb}
\usepackage{xcolor}

\usepackage{amsmath}

\makeatletter
\def\ps@IEEEtitlepagestyle{%
  \def\@oddfoot{\mycopyrightnotice}%
  \def\@evenfoot{}%
}
\def\mycopyrightnotice{%
  {\footnotesize 978-1-5386-6808-5/18/\$31.00 \copyright2018 IEEE \hfill}% <--- Change here
  \gdef\mycopyrightnotice{}% just in case
}

\ifCLASSINFOpdf
\else
\fi
\begin{document}

% \IEEEoverridecommandlockouts
% \IEEEpubid{\makebox[\columnwidth]{978-1-5386-6808-5/18/\$31.00 \copyright 2018 IEEE \hfill }} 
% \hspace{\columnsep}\makebox[\columnwidth]{\hfill }

% \def\mycopyrightnotice{%
%   {\footnotesize The copyright belongs to me!\hfill}% <--- Change here
%   \gdef\mycopyrightnotice{}% just in case
% }

\title{U-PoT: A Honeypot Framework for UPnP-Based IoT Devices}
%\thispagestyle{plain}
%\pagestyle{plain}

% \author{\IEEEauthorblockN{Muhammad Azizul Hakim}
% \IEEEauthorblockA{Florida International University\\
% Miami, Florida \\
% Email: mhaki005@fiu.edu}
% \and
% \IEEEauthorblockN{Hidayet Aksu}
% \IEEEauthorblockA{Florida International University\\
% Miami, Florida\\
% Email: haksu@fiu.edu}
% \and
% \IEEEauthorblockN{A. Selcuk Uluagac}
% \IEEEauthorblockA{Florida International University\\
% Miami, Florida \\
% Email: suluagac@fiu.edu}
% \and
% \IEEEauthorblockN{Kemal Akkaya}
% \IEEEauthorblockA{Florida International University\\
% Miami, Florida \\
% Email: kakkaya@fiu.edu}}

% \author{
%     \IEEEauthorblockN{Muhammad Azizul Hakim\IEEEauthorrefmark{1}, Hidayet Aksu\IEEEauthorrefmark{1}, A. Selcuk Uluagac\IEEEauthorrefmark{1}, Kemal Akkaya\IEEEauthorrefmark{1}}
%     \IEEEauthorblockA{\IEEEauthorrefmark{1}Florida International University
%     \\\{mhaki005, haksu, suluagac, kakkaya\}@fiu.edu}
%     \IEEEauthorblockA{\IEEEauthorrefmark{2}Institution2
%     \\\{2, 3\}@def.com}
% }

% \author{
%     \IEEEauthorblockN{Muhammad Azizul Hakim, Hidayet Aksu, A. Selcuk Uluagac, Kemal Akkaya}
%     \IEEEauthorblockA{CSL Lab\\\ 
%     Florida International University
%     \\\{mhaki005, haksu, suluagac, kakkaya\}@fiu.edu}
    
% }  , and Kemal Akkaya

\author{\IEEEauthorblockN{Muhammad A. Hakim, Hidayet Aksu, A. Selcuk Uluagac, Kemal Akkaya}
\IEEEauthorblockA{ Cyber-Physical Systems Security Lab (CSL) \\ 
Department of Electrical and Computer Engineering \\
Florida International University \\
\{mhaki005,haksu,suluagac,kakkaya\}@fiu.edu}

}


%Telephone: (800) 555--1212, Fax: (888) 555--1212}


\maketitle

\begin{abstract}
%\boldmath
%\blindtext[1]
The ubiquitous nature of the IoT devices has brought serious security implications to its users. A lot of consumer IoT devices have little to no security implementation at all, thus risking user's privacy and making them target of mass cyber-attacks. Indeed, recent outbreak of Mirai botnet and its variants have already proved the lack of security on the IoT world. Hence, it is important to understand the security issues and attack vectors in the IoT domain. Though significant research has been done to secure traditional computing systems, little focus was given to the IoT realm. In this work, we reduce this gap by developing a honeypot framework for IoT devices. Specifically, we introduce U-PoT: a novel honeypot framework for capturing attacks on IoT devices that use Universal Plug and Play (UPnP) protocol. A myriad of smart home devices including smart switches, smart bulbs, surveillance cameras, smart hubs, etc. uses the UPnP protocol. Indeed, a simple search on Shodan IoT search engine lists 1,676,591 UPnP devices that are exposed to public network. The popularity and ubiquitous nature of UPnP-based IoT device necessitates a full-fledged IoT honeypot system for UPnP devices. Our novel framework automatically creates a honeypot from UPnP device description documents and is extendable to any device types or vendors that use UPnP for communication. To the best of our knowledge, this is the first work towards a flexible and configurable honeypot framework for UPnP-based IoT devices. We released U-PoT under an open source license for further research on IoT security and created a database of UPnP device descriptions. We also evaluated our framework on two emulated deices. Our experiments show that the emulated devices are able to mimic the behavior of a real IoT device and trick vendor-provided device management applications or popular IoT search engines while having minimal performance ovherhead.

%to think that the emulated devices are some real physical devices thus capturing incoming traffic from real vendor-provided device management applications (e.g., from smartphones). The performance evaluation shows that U-PoT can be used to deploy honeypots at a large scale in a single virtual machine with minimum overhead.

\end{abstract}


\begin{IEEEkeywords}
iot honeypot, threat detection, system security.
\end{IEEEkeywords}

% \fi
\IEEEpeerreviewmaketitle

\section{Introduction}
In recent years, the use of IoT devices has increased to a great extent. Increased application of smart wearables, home improvement systems etc. connected more users to the Internet. Indeed, by 2020, Internet connected devices will rapidly grow from 5 billion to 24 billion\cite{luo2017iotcandyjar}. Securing this huge network of connected devices has already become a prime concern due to several existing attacks of botnets like Mirai\cite{antonakakis2017understanding} and its variants Qbot\cite{muncaster_2014}, Okiru\cite{okiru}, etc. Most of the IoT devices are wireless connected making them an easy target of eavesdropping. Having low computing resources, these devices cannot implement a complex security scheme thus making them a prime target of malicious users. Furthermore, not all users are well-versed about security issues and tend to keep the default usernames and passwords for their devices, thus making an attack easier for a hacker. In fact, this was the main reason how the Mirai botnet\cite{antonakakis2017understanding} was powered by a list of 62 default usernames and passwords and was able to bring down some of the popular websites and major part of the Internet infrastructure. 

In the IoT ecosystem, a significant number of the IoT devices use Universal Plug and Play (UPnP) protocol to communicate. According to a research\cite{moore2013security}, some 6900 network-aware products from 1500 companies at 81 million IP-addresses responded to their UPnP discovery requests. 20\% of those 81 million systems also exposed a Simple Object Access Protocol (SOAP) API to the Internet providing an easy entry point to malicious attackers. The same research has also pointed out many vulnerabilities in the UPnP device SDK. Although some of those vulnerabilities are patched, some are still at large and with a new patch, there is chance to add new vulnerabilities. Another UPnP scanner built using ZMap\cite{durumeric2013zmap} network scanner found  3.4 million UPnP devices with known vulnerabilities. As of this writing, Shodan IoT search engine has listed around 1,676,591 publicly exposed UPnP devices. Therefore, it is very important to develop a cost-efficient method to identify vulnerabilities in UPnP-based IoT devices. Honeypots are very promising in this direction as they have been proven useful\cite{brzeczko2014active} for disclosing and analyzing these vulnerabilities without exposing critical assets.

Today, there are so many honeypot implementation for traditional computing systems, but very few for IoT devices and none for the UPnP devices. Due to heterogeneous nature of IoT devices, it is very difficult to have an IoT honeypot that covers a wide range of devices. In this paper, we solved this problem by emulating the entire IoT platform. We introduce a novel honeypot framework called U-PoT to create emulated honeypot device for UPnP-based IoT devices. U-PoT is agnostic of device type or vendor, flexible and easily configurable for any UPnP-based devices.

% There are so many honeypot implementation for traditional computing system, but very few for IoT devices and none for the UPnP devices. Moreover, most of the IoT honeypots are not full scale and only simulates a Telnet, SSH or few other protocols which do not really capture the real characteristics of IoT devices. They are not as interactive as a real device making it very easy for the attacker to find out that it is a honeypot and thus failing to capture real attacker behavior. In this paper, our goal is to develop an interactive honeypot for UPnP devices to capture IoT attacks.

% User interaction is important for any honeypot to capture useful information from the attacker. Most of the IoT honeypots are not interactive. They fail to engage user by providing an interactive session. Most of them simulate only some protocol or network stack. An intelligent attacker can easily find out their limitations and may lose interest. For example, an attacker targeting a smart switch will try to get a telnet session to the switch. It is on the best interest of the attacker to hide his intentions. So before doing anything, an intelligent attacker will try to make sure that he is communicating with a real physical switch, not some honeypot. Most of the conventional honeypots will fail to pass this check. We solved this problem by emulating the entire IoT platform.

\noindent \textbf{\textit{Contributions}:} In this work, we focus on the development of an interactive emulated IoT device that uses UPnP protocol. Utilizing the device description document of the UPnP devices, we developed an open-source, flexible, configurable, and interactive honeypot platform for UPnP-based IoT devices. In summary, our contributions include:

\begin{itemize}
    \item a novel framework to automatically create an emulated UPnP device from a UPnP device description document.
    \item an emulated version of a smart switch that uses UPnP protocol for communication.
    \item evaluation of our framework in different test scenarios.
    \item creation of an open source database of UPnP descriptions for variety of devices.
    \item and finally, in support of open science, the U-PoT framework is made available\footnotemark to research community under an open-source license. %For more information, please see https://github.com/azizulhakim/u-pot/.
\end{itemize}

\footnotetext[1]{U-PoT project for download: https://github.com/azizulhakim/u-pot/}

\noindent \textbf{\textit{Organization}:} The rest of the paper is organized as follows: Section II discusses the related work. Section III gives some background on honeypots; Section IV explains UPnP device architecture. U-PoT famework architectural design is explained in Section V. Section VI shows experimental setup and evaluation and we conclude the paper in Section VII.

\section{Related Work}
Research on honeypot technologies has been introduced at the end of 1990s and since then, it has evolved. Significant research work has been put to resolve new security threats and vulnerabilities. Honeyd\cite{provos2004virtual} is a popular virtual honeypot framework to simulate computer systems at the network level. It simulates the TCP/IP stack of an operating system and supports TCP, UDP, and ICMP protocols. To match the network behavior of the configured operating system, honeyd's personality engine modifies the response packet before sending a response for incoming requests. Nepenthes\cite{baecher2006nepenthes} honeypot platform emulates only the vulnerable parts of a service. 

Though significant research was done on honeypot technologies for traditional computing systems, there is little work done for IoT honeypots. To the best of our knowledge, the first IoT honeypot was introduced by IoTPOT\cite{pa2015iotpot}, a low-interaction honeypot for Telnet protocol. Other Telnet and SSH based honeypots\cite{vsemic2017iot,pauna2014casshh, wagener2011adaptive} are also common for the IoT domain. However, these honeypots are low-interaction and do not represent an actual IoT device behavior. They simulate only a part of the network sub-system and face difficulty in engaging user in an interactive session to capture useful information. ThingPot\cite{wang2017thingpot} is the first low-interaction honeypot to emulate the Philips Hue smart lighting system, but it is not easily extendable to other devices. SIPHON\cite{guarnizo2017siphon} solved the problem of low-interaction honeypot by deploying 80 high-interactive devices with a diverse set of IPs located in different geographical areas by using only seven real IoT devices. This setup helps to engage users for longer sessions, but it is expensive and the cost increases with increased number of physical devices. A new concept of intelligent interaction honeypot was introduced by IoTCandyJar\cite{luo2017iotcandyjar} which emulates the request-response pattern of an IoT device. It collects a seed database of IoT request extracted using a low-interaction honeypot and probes online IoT devices to receive response for those requests. It uses machine learning to deduce response for unknown requests and update the database with newly deduced knowledge. However, this approach is highly dependent on the initial seed request. So a malicious attacker can create fake devices and deploy large number of those fake devices to manipulate the operation of it. All these honeypots are server honeypots that expose server services and wait to be attacked. On the other hand, client honeypots\cite{seifert2007honeyc, nazario2009phoneyc, akiyama2010design, alosefer2010honeyware} are different types of honeypots to detect malicious servers. In this work, we keep our focus on server honeypots only. We developed a novel framework for automatically creating interactive honeypot for IoT devices that use UPnP as their communication protocol. 

\noindent \textit{\textbf{Difference from the existing work: }} U-PoT solves the problem of existing low-interaction honeypots while providing the advantages of high-interaction honeypots. It is agnostic of device type or vendor, flexible, and easily configurable for any UPnP-based devices. Operation of U-PoT is not dependent on any initial seed requests and hence, it solves the problem possessed by IoTCandyJar. To the best of our knowledge, this is the first work to focus on the interactive nature of IoT devices to provide a cost-effective, flexible and configurable honeypot framework for UPnP-based IoT devices.

\section{Background}
Honeypot is a very instrumental tool widely used in identifying and analyzing unknown attack vectors. Generally, it is deployed in a controlled environment to attract attackers. From the attacker's perspective, a honeypot appears just like any regular system, but in reality, the interactions between the attacker and the honeypot are closely monitored and collected for further analysis by researchers. This helps researchers to discover new vulnerabilities in a system and identify appropriate mitigation techniques.

Honeypots are traditionally classified into two categories based on their interaction level: low-interaction and high-interaction. \textit{Low-Interaction Honeypots:} A low-interaction honeypot is an emulated service that gives attackers very limited level of interaction. Most of the time, these honeypots are just a collection of network service implementations like Telnet, SSH etc. They are lightweight by design and can be developed in less time and deployed in large scale. Because of the limited interaction provided by them, they can collect only limited information and are easily identifiable by an intelligent attacker. \textit{High-Interaction Honeypots:} A high-interaction honeypot tends to provide a full-fledged system. An attacker can interact with a real system and has access to all system functionality that a normal user has access to. They are not easily detectable and can collect more information about the attacker. They are complex and costly to implement and deploy.

Low-interaction honeypots only support a few features of a system rather than supporting the entire system behavior. They are not interactive and fail to engage users for a longer session. Vulnerabilities on IoT devices are usually highly dependent on specific device brand or even firmware version. This leads to the fact that attackers tend to perform several checks on the remote host to gather more device information before launching the exploit-code. Limited level of interaction provided by a low-interaction honeypot is not enough to pass the check and fail to capture real attack \cite{luo2017iotcandyjar}. This problem can be solved by deploying high-interactive IoT devices, but that would be costly. In fact, there are medium-interaction honeypots that are built with the help of full system virtualization. In this approach, a researcher takes the system image for a target system and deploys it in a virtual environment to emulate the real behavior of a physical device. Unfortunately, access to system images for popular IoT devices are limited. 
%It is possible to extract the image from the device using a debugger, but it requires a lot of technical effort for a novice user, it is error prone, and not scalable. Furthermore, many vendors apply anti-debugging techniques on their devices making this technique useless. 
In this paper, we overcome these problems by introducing an emulated IoT device (i.e., smart bulb) and showed how it can be used as a replacement for a full system virtualization-based honeypot.

In our system, we selected IoT devices that use UPnP protocol for communication. The rationale for this selection stems from the fact that UPnP is widely deployed on a lot of IoT devices and enormous discussion has been made by researchers about its security issues. Having a honeypot tool targeted to UPnP device class will help researchers to deploy the same tool for a large variants of devices and vendors to fight against vulnerabilities in the IoT domain.

\section{UPnP Device Architecture}
In this section, we present a short description on the architecture of a UPnP device and their communication mechanisms as this will help to understand the implementation of the U-PoT framework. Universal Plug and Play (UPnP) device architecture\cite{upnp_architecture} is an extension to plug and play technology for networked devices. %It is a set of networking protocols that permit networked devices to seamlessly discover each other's presence on the network and establish functional network services. 
UPnP technology leverages Internet protocols such as IP, TCP, UDP, HTTP and XML to support zero-configuration, "invisible" networking, and automatic discovery for a breadth of device categories from a wide range of vendors. Every UPnP device is defined by an XML-based device schema for the purpose of enabling device-to-device interoperability in a scalable, networked environment. 

UPnP devices can be classified in two categories: \textit{controlled devices} and \textit{control points}. Controlled devices act as a server which is responsible for delivering a service. For example, a smart switch is a controlled device that implements some service and exposes them to home users to on/off the switch, set rules, update configuration etc. Generally a home user takes advantage of those services using a mobile application (e.g., a smartphone app) which is referred to as a \textit{control point} in the UPnP context. Furthermore, UPnP stack consists of the following layers.

\subsection{Discovery Layer}
Discovery is the first step in UPnP networking. The discovery protocol allows a device to advertise its service to other devices on the network or allow other devices to search for an UPnP device on the network. The discovery message contains some essential specifics of the device including a pointer to more detailed information of the device. When a new device joins the network, it may multicast a discovery message searching for an interesting device. An UPnP device must listen to the standard multicast address and must respond if the search criteria in the discovery message matches with itself. UPnP devices use Simple Service Discovery Protocols (SSDP) for device discovery or advertisement. When a control point (e.g., a smartphone app) desires to search the network for a device, it sends a multicast request with \textit{M-SEARCH} method on the reserved address and port (239.255.255.250:1900). The format of the \textit{M-SEARCH} message is shown in Figure \ref{fig:m-search-request-format}. An UPnP device responds to a discovery message with some basic information of the device which includes UPnP type, universally-unique identifier, and a URL to the device's UPnP description.

\begin{verbatim}
M-SEARCH * HTTP/1.1
HOST: 239.255.255.250:1900
MAN: "ssdp:discover"
MX: seconds to delay response
ST: search target
USER-AGENT: optional field
\end{verbatim}
\begin{figure}[!h]
\caption{\textit{M-SEARCH} Request Format.}
\label{fig:m-search-request-format}
\end{figure}

\begin{figure}
\includegraphics[width=0.95\linewidth]{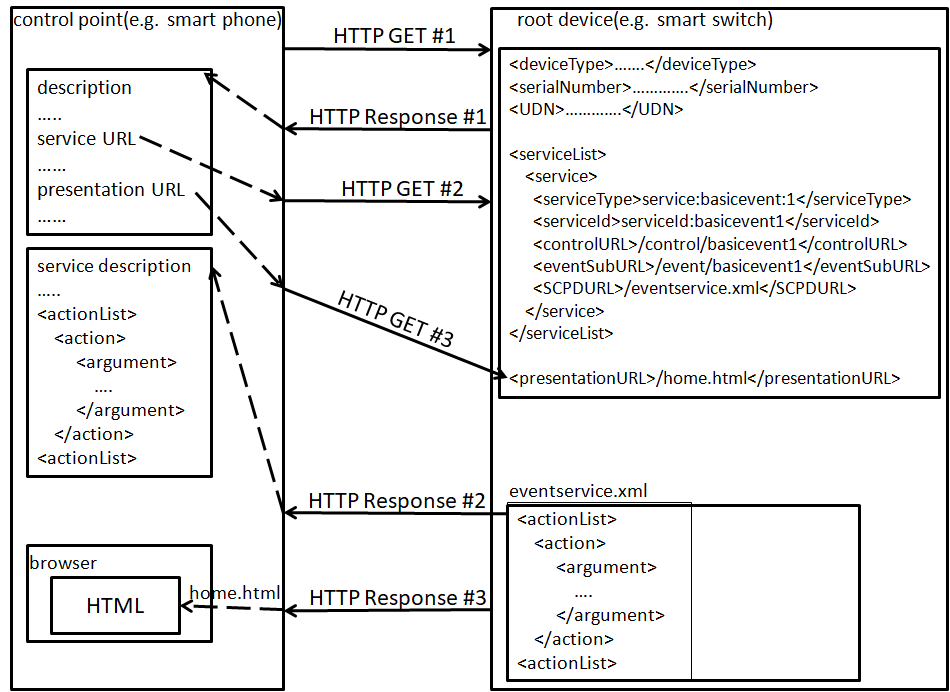}
%\includesvg[width=1.0\linewidth]{figures/fig_description.svg}
\caption{UPnP Description Architecture.}
\centering
\label{fig:fig_description}
\end{figure}

\subsection{Description Layer} \label{description_layer}
After discovering a device, the control point has very little information about the device. To interact with the device, the control point has to retrieve a description of the device and its capabilities from the URL provided by the device in the discovery message. This UPnP device description is an XML formatted data partitioned into two logical parts: a device description describing the physical and logical containers, and service descriptions describing the capabilities exposed by the device. Figure \ref{fig:fig_description} shows the description architecture of a UPnP device. The description includes vendor specific information. For each service in the device, the device description lists the service type, service name, a URL for the service description, a URL for control, and a URL for eventing. 

The UPnP description for a service defines actions that are accessible by a control point and their arguments. It also defines a list of state variables and their data type, range, and event characteristics. The state variable represents device state in a given time. Each service associates with one or more action. Each action has input and output arguments. Each argument corresponds to one of the state variables. A single physical device may include multiple logical devices. The device description also includes the description of all the embedded devices and one or more URL for presentation. The control point issues \textit{HTTP GET} request to get a specific URL.
\subsection{Control \& Eventing Layers}
Once the control point has information about the UPnP device and its services, it can invoke actions from those services and receive a response. To invoke an action, the control panel sends a control message to the fully qualified control URL for the service. The service returns result or error in response. Events are published to all interested control points if the effect of the action makes change to the state variable of that service.

\subsection{Presentation Layer}
If a device has a URL for presentation, then the control point can retrieve a page from this URL, load the page into a browser and depending on the capabilities of the page, allow a user to control the device and/or view device status.

\section{U-PoT Design and Implementation}
\begin{figure}[!h]
\includegraphics[width=1.0\linewidth]{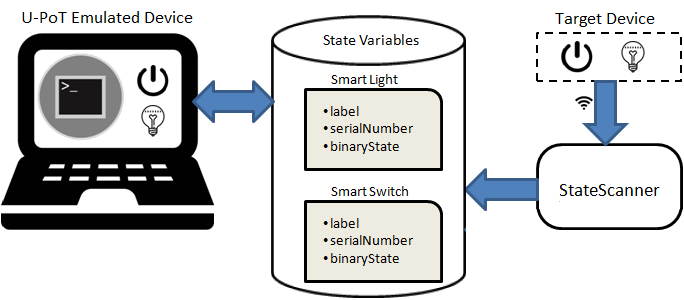}
\caption{U-PoT system architecture. U-PoT is availabe for download at https://github.com/azizulhakim/u-pot as an open-source project}
\centering
\label{fig:system_architecture}
\end{figure}

In this section, we present the implementation of the U-PoT framework. A schematic view of the system architecture is shown in Figure \ref{fig:system_architecture}. We start by outlining our choice of IoT devices and the protocols they use and then present the proof-of-concept design of the system.

% \begin{figure}
% \includegraphics[width=1.0\linewidth]{architecture.png}
% \caption{U-PoT system architecture}
% \centering
% \label{fig:system_architecture}
% \end{figure}

\subsection{Target Device}
In this sub-section we introduce our selection of target device, a real UPnP-based IoT device whose device description is used for the implementation of the U-PoT framework. For the implementation of U-PoT, we selected Belkin Wemo smart switch \cite{belkin}. The rationale for this selection stems from the fact that, smart switches are the most common IoT devices and are widely used in smart home settings. Before going into implementation details, we introduce a short description about the working principles of the device.

Belkin Wemo Smart Switch is a UPnP protocol-based device that uses SOAP messaging specification for exchanging structured information. It has 12 services implemented in its firmware. Each service has one or more action for control and event subscription. Using a control application, a user can invoke an action from these services to manipulate device states, add rules, update firmware, etc. It also hosts some image and HTML files for presentation. The control point is an Android or iOS application that uses SSDP discovery request to discover the device on the network. When the application starts, it issues \textit{M-SEARCH} command on the multicast address 239.255.255.250:1900. Figure \ref{fig:m-search} shows the attributes of the \textit{M-SEARCH} command issued by U-PoT to discover target device's information. As noted earlier, any UPnP device listening to this multicast address will respond to the \textit{M-SEARCH} request with its information. In the U-PoT's case, the target device responds with its UPnP type, unique identifier, and a URL to its UPnP description XML file as in Figure \ref{fig:m-search-response}. The control application parses the XML file to extract more information about the device and displays the relevant information.

\begin{verbatim}
M-SEARCH * HTTP/1.1
HOST: 239.255.255.250:1900
MAN: "ssdp:discover"
MX: 2
ST: upnp:rootdevice
\end{verbatim}
\begin{figure}[!h]
\caption{\textit{M-SEARCH} Request for Target Device.}
\label{fig:m-search}
\end{figure}

\begin{Verbatim}[commandchars=\\\{\}]
(('10.0.0.11',49153),'HTTP/1.1 200 OK
\textcolor{red}{Location: 10.0.0.11:49153/setup.xml}
Cache-Control: max-age=1800
Server: Unspecified, UPnP/1.0
uuid:XXXXXXXXXXXXXX::upnp:rootdevice')
\end{Verbatim}
\begin{figure}[!h]
\caption{\textit{M-SEARCH} Response from WeMo Switch.}
\label{fig:m-search-response}
\end{figure}

% \begin{verbatim}
% M-SEARCH * HTTP/1.1
% HOST: 239.255.255.250:1900
% MAN: "ssdp:discover"
% MX: 2
% ST: upnp:rootdevice
% \end{verbatim}
% \begin{figure}[!h]
% \caption{\textit{M-SEARCH} Request for Target Device.}
% \label{fig:m-search}
% \end{figure}

%\begin{figure}
%\centering
%\begin{BVerbatim}
%M-SEARCH * HTTP/1.1
%HOST: 239.255.255.250:1900
%MAN: "ssdp:discover"
%MX: 2
%ST: upnp:rootdevice
%\end{BVerbatim}
%\caption{C++ code}
%\end{figure}

\subsection{State Scanner}
U-PoT state scanner is responsible for extracting the description layer of a UPnP enabled device. UPnP description layer is an XML definition of the device that describes the device capabilities. It contains device configuration information and different service descriptions. Please refer to \ref{description_layer} for more information about the description layer of an UPnP device. The state scanner issues a multicast M-SEARCH command on 239.255.255.250:1900 to discover available UPnP devices on the network. The response to M-SEARCH request contains device UUID and an URL to the device description file as shown in Figure \ref{fig:m-search-response}. The URL contains the UPnP root description XML and works as an entry point to the description layer of the device. Once state scanner retrieves the entry point, the remaining work is done in two steps.

\noindent \textbf{\textit{Crawling:}}
A UPnP device might host multiple service descriptor XML files defining the \textit{Control \& Eventing Layer} and some image or HTML file for the \textit{Presentation Layer}. U-PoT state scanner parses the root description to extract the URLs to those resources. Once the paths are extracted, it starts crawling those URLs using \textit{HTTP Get} method and store them in local file system. The crawled files are later hosted from the U-PoT emulated device. Each UPnP service descriptor defines multiple actions and action arguments that are accessible by a control point. It also defines a list of state variable which represents the device state in a given time. The crawler parses the service descriptor XML to generate a curated list of state variables, actions and action arguments. This list is later used by the U-PoT emulated device to represent the runtime state of the emulated device.

\noindent \textbf{\textit{Scanning:}}
After crawling all description and presentation layer information, U-PoT enters into the scanning state. During the scanning state, U-PoT tries to create a snapshot of a valid start point for the emulated device. Having a valid start point is important for the emulated device as without it the emulated device might fail to establish the initial handshaking communication with a genuine control point. To create a valid snapshot, U-PoT invokes the actions from each service it extracted during the crawling step. From the device context, an action is a function that can be invoked by a client to set the value of a state variable or get the value of a state variable. The U-PoT state scanner gets the value of a state variable by formulating a SOAP request which it uses to invoke a \textit{get} action and retrieves its value from the response. Finally, it updates the state variable database using the value received. This database represents a valid snapshot of our target device and can be used to initialize an emulated U-PoT device.

% UPnP service descriptors defines actions and action arguments that are accessible by a control point. It also defines a list of state variable which represents the device state in a given time. In scanning state, U-PoT make use of the crawled service descriptors to list all action, action arguments and state variables. U-PoT state scanner parses the service description to discover state variables specific for a service and adds them to the state variable database. To get the run-time value of a state variable, U-PoT state scanner makes use of the actions provided by the service. From the device context, an action is a function that can be invoked by a client to set the value of a state variable or get the value of a state variable. The U-PoT state scanner creates a SOAP request to make a call to the \textit{get} actions and retrieves its value from the response.  Finally, it updates the database using the value received. We used this database to initialize the U-PoT device to an initial state that represents a state of the physical IoT device. 

The IoT device and the service descriptions are written by device vendors and is usually based on a standard UPnP Device/Service Template. As the device and service descriptions are created from a generic template and all UPnP device vendors follow these templates to create the descriptions for their device, U-PoT state scanner is vendor-agnostic and usable to automatically extract information from any UPnP device from any vendor.

\begin{figure}[!h]
\includegraphics[width=1.0\linewidth]{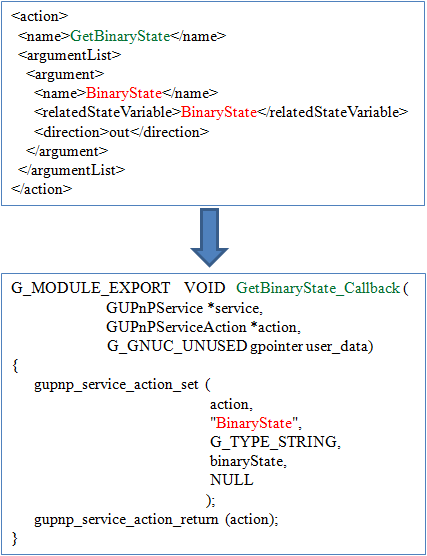}
\caption{Generation of U-PoT Device Callback from Action.}
\centering
\label{fig:action_callback}
\end{figure}

\subsection{U-PoT Device}
The last step to build the U-PoT honeypot framework is to create the emulated device that can listen to incoming requests and return/update its state accordingly. Similar to a regular UPnP device, the U-PoT device should work on two different modes. In the discovery mode, U-PoT devices should respond with device information and device state that is extracted using the scanner. In the normal operation mode, they should respond to incoming requests to change device state or notify device state. The U-PoT framework achieves this by using the \textit{gupnp}\cite{gupnp} library from the GNOME project. This library handles the initialization of a UPnP device, hosting device description files and other static resources, responding to SSDP discovery requests, and serialize/deserialize SOAP requests. Initially, U-PoT generates \textit{gupnp} compatible code for a skeleton UPnP device and initializes service and presentation layer using the service and presentation descriptor XMLs extracted during \textit{scanning} state. Next the U-PoT framework adds action handlers for different actions, that could be invoked by a control point. An action can be a request to read a state variable or a request to write new value to a state variable. For example, a control point might want to know whether a smart switch is turned on/off or might want to turn on/off a switch. Trying to turn on/off the switch invokes an action to write new value on the corresponding state variable. Similarly, trying to know whether a switch is on/off invokes an action to read corresponding state variable. U-PoT automatically generates an empty function block for each action. Next it adds body of the function by using the input-output parameters extracted by \textit{State Scanner} for each action. If the action corresponds to a state read request, U-PoT will reference the state variable to retrieve the value of the variable and create SOAP response. Figure \ref{fig:action_callback} shows how U-PoT maps an UPnP action to U-PoT callback function for a read request. On the other hand, for a write request, U-PoT extracts the value of the incoming state variables from the incoming SOAP message and updates the corresponding variables in state variable database. This U-PoT generated code is compatible with \textit{gupnp} library and can be compiled and linked with \textit{gupnp} to create an executable for the U-PoT emulated device. Automating the code generation for device actions makes the framework device type/vendor agnostic and one can emulate any UPnP device as long as they have access to the device/service description files. %The open source code of U-PoT can be freely downloaded from https://github.com/azizulhakim/u-pot/.

\begin{figure}[bh!]
\centering
\includegraphics[width=0.55\linewidth]{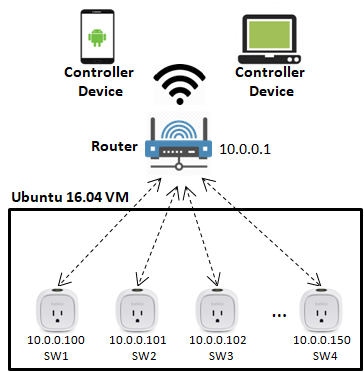}
\caption{Multiple U-PoT emulated device (smart switches) setup in a single virtual machine.}
\centering
\label{fig:fig_deployment}
\end{figure}

\begin{figure}[bh!]%
\centering
\subfigure[][]{%
\label{fig:fig_response-a}%
\includegraphics[width=0.75\linewidth]{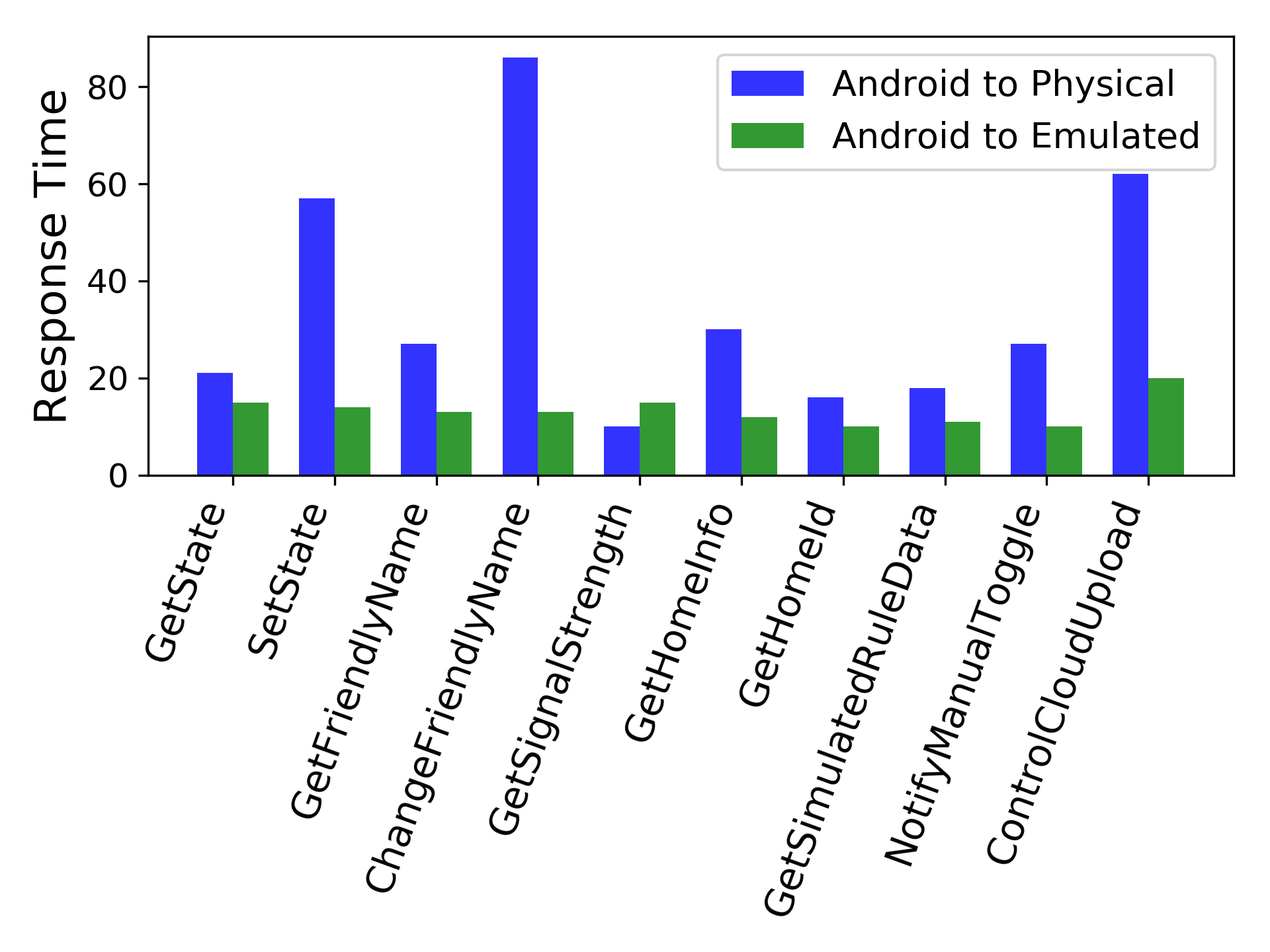}}%

\subfigure[][]{%
\label{fig:fig_response-b}%
\includegraphics[width=0.75\linewidth]{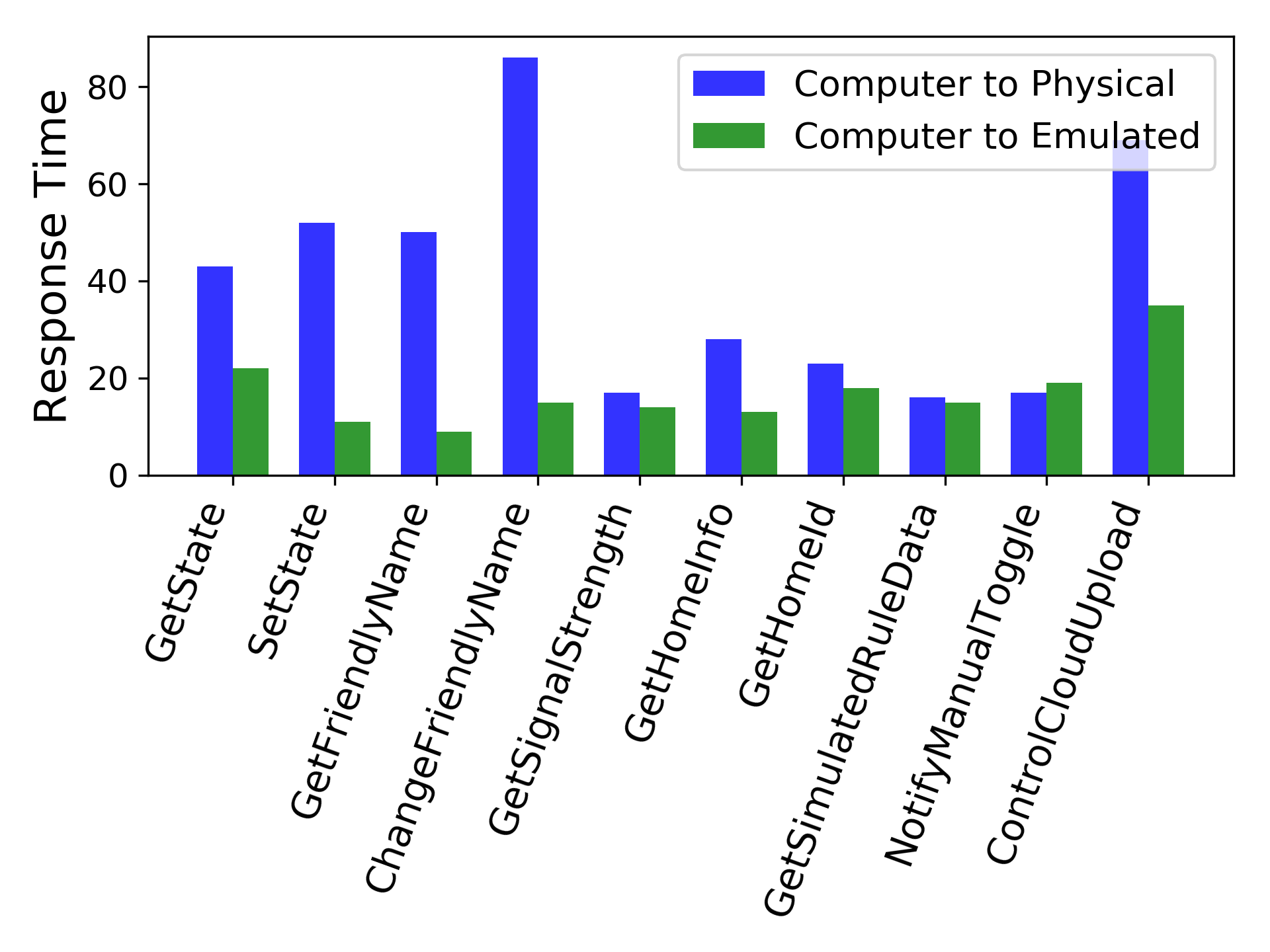}} \\

\caption[A set of four subfigures.]{Comparison of response time between the physical and emulated devices:
\subref{fig:fig_response-a} requests are sent from an Android based control device; and,
\subref{fig:fig_response-b} requests are sent from the Computer.}%
\label{fig:fig_response}%
\end{figure}

\begin{figure*}[h!]%
\centering
\subfigure[][]{%
\label{fig:fig_scalibility-a}%
\includegraphics[width=0.45\linewidth]{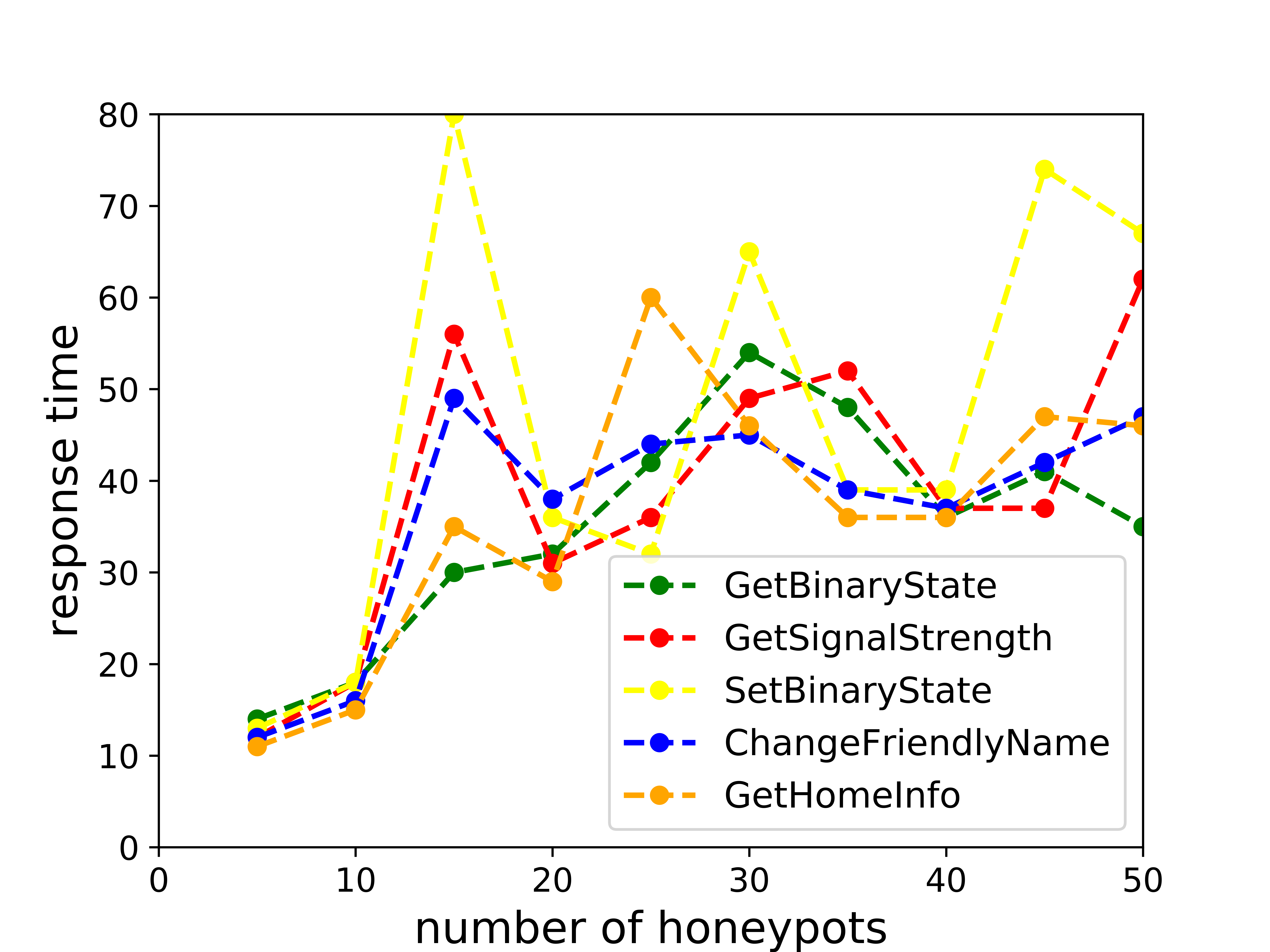}}%
\subfigure[][]{%
\label{fig:fig_scalibility-b}%
\includegraphics[width=0.45\linewidth]{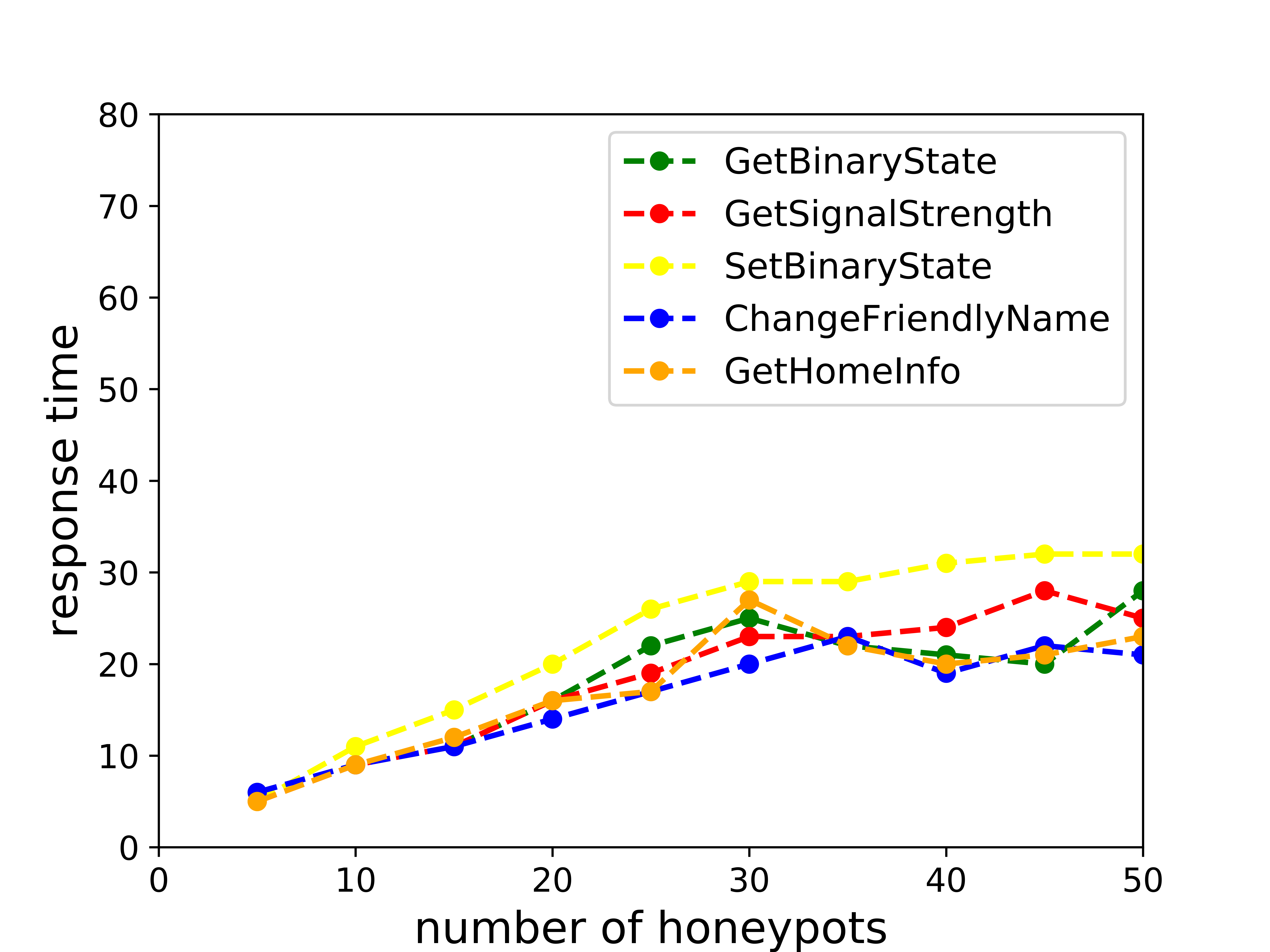}} \\
\subfigure[][]{%
\label{fig:fig_scalibility-c}%
\includegraphics[width=0.45\linewidth]{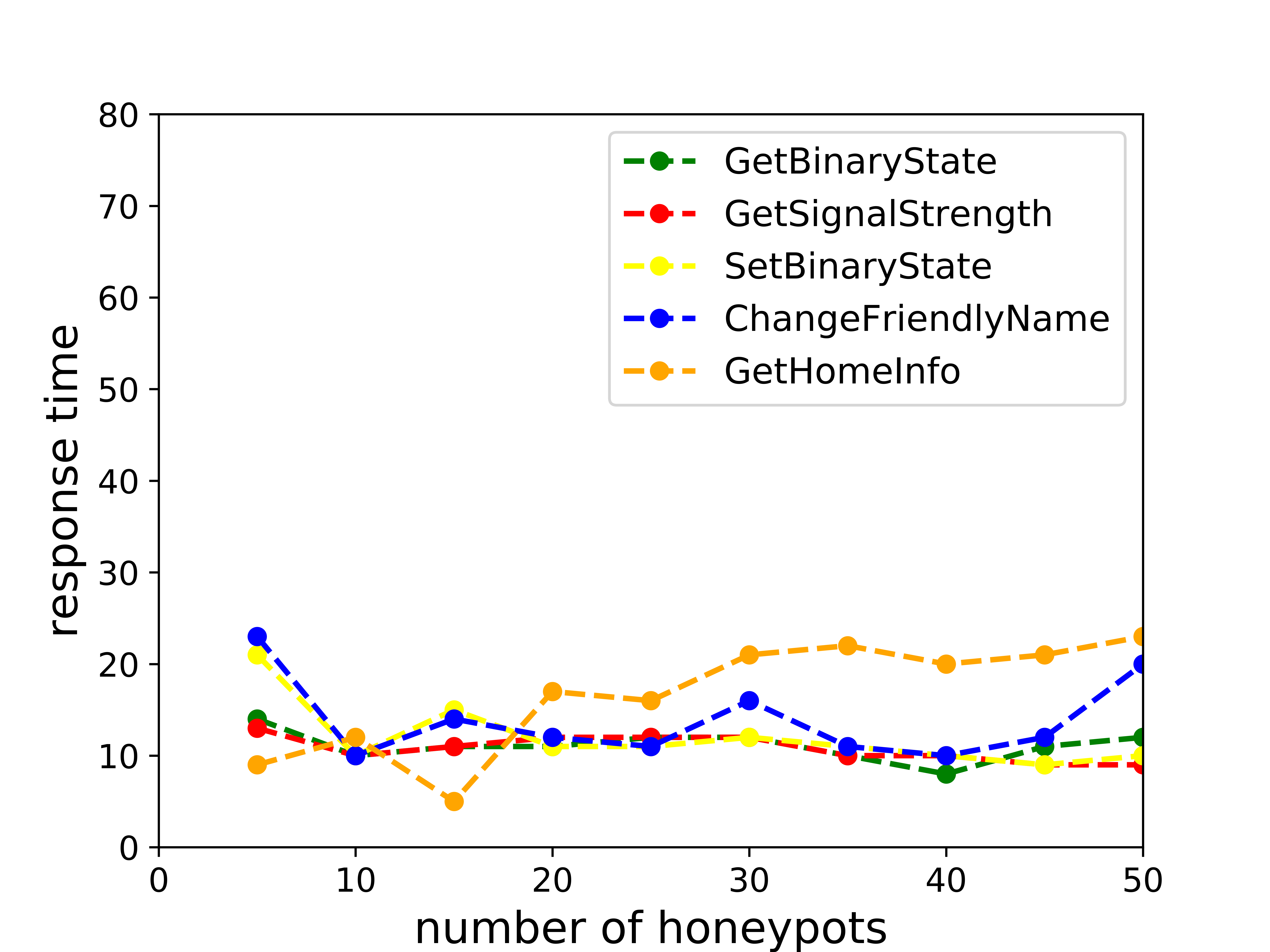}}%
\subfigure[][]{%
\label{fig:fig_scalibility-d}%
\includegraphics[width=0.45\linewidth]{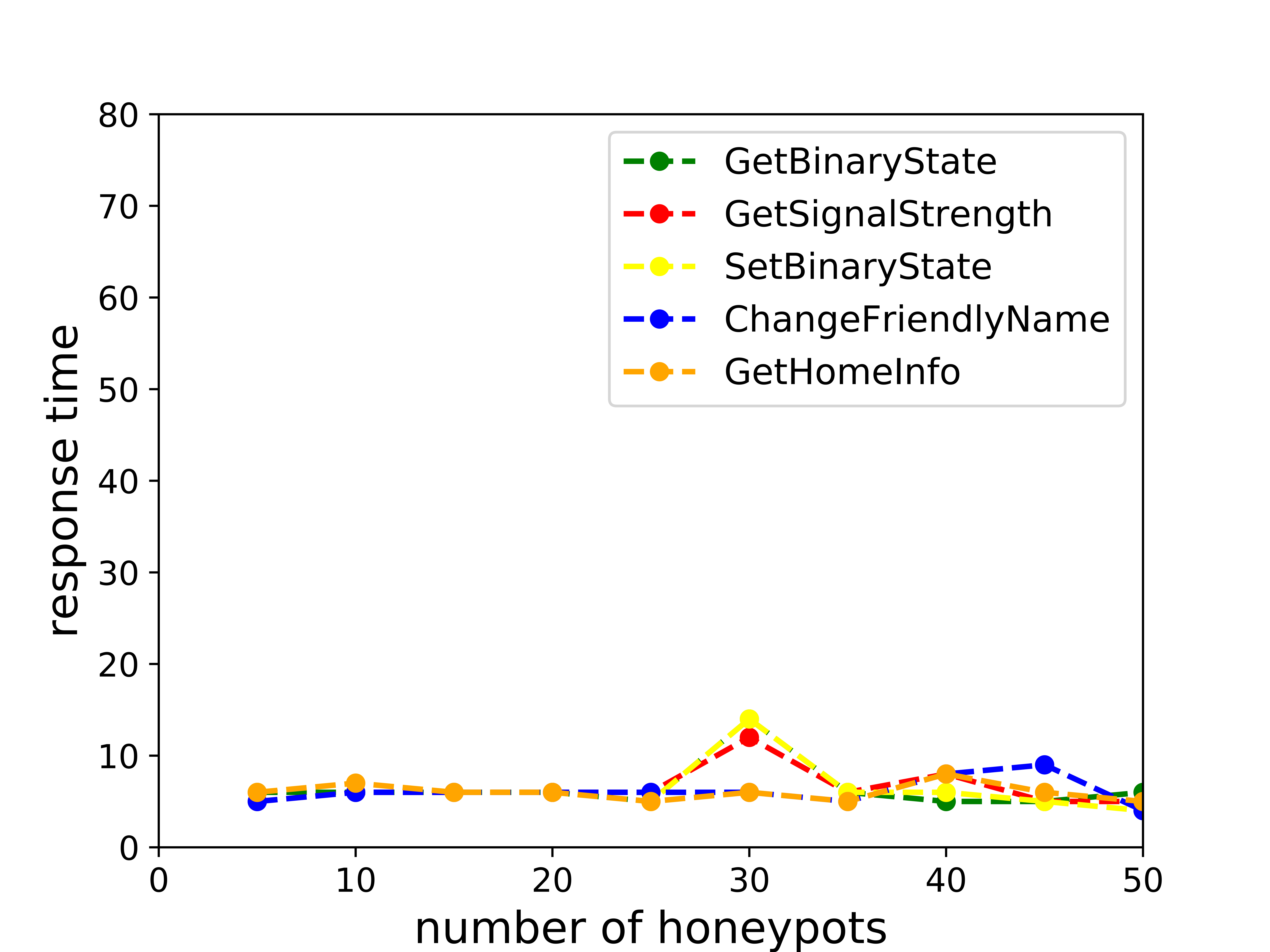}}%
\caption[A set of four subfigures.]{Change in response time with increased number of honeypots:
\subref{fig:fig_scalibility-a} Android-to-emulated device communication for concurrent requests to all the emulated IoT devices;
\subref{fig:fig_scalibility-b} computer-to-emulated device communication for concurrent requests to all the emulated IoT devices;
\subref{fig:fig_scalibility-c} Android-to-emulated device communication for concurrent requests to a subset (5) of the emulated IoT devices; and,
\subref{fig:fig_scalibility-d} computer-to-emulated device communication for concurrent requests to a subset (5) of the emulated IoT devices.}%
\label{fig:fig_scalibility}%
\vspace{-10pt}
\end{figure*}

\section{Performance Evaluation}
In this section, we evaluate the performance of the U-PoT framework. We use U-PoT's ability to create an emulated device from unknown UPnP device description as the primary evaluation criteria. We also make a comparative analysis of the response time between a real physical device and the U-PoT emulated device. Finally, we perform scalability analysis to evaluate the average response time of the emulated device under varying number of device deployments. With these, in the evaluation, our goal was to answer the following questions:

\begin{itemize}
    \item Can a U-PoT emulate a real physical device and fool a real device control/management application (i.e., the accompanying vendor-provided Smartphone App)?
    \item Can a U-PoT emulated device serve client requests within a reasonable response time?
    \item Can the U-PoT framework deploy large number of emulated devices with minimal overhead (i.e., scalability)?
\end{itemize}

\subsection{Emulate a Physical Device}
We evaluate U-PoT's ability to mimic a real physical device with the help of a vendor provided device management application and the \textbf{open} \textbf{H}ome \textbf{A}utomation \textbf{B}us (openHAB)\cite{openHAB} framework. openHAB is a very popular, open-source, technology-agnostic home automation platform. It integrates different home automation systems, devices, and technologies into a single solution. It provides uniform user interfaces, and a common approach to automation rules across the entire system, regardless of the number of manufacturers and sub-systems involved. The basic building blocks of openHAB are many plugins for different types of devices from different manufacturers. The plugins are called \textit{binding}. For our evaluation, we installed openHAB with two different bindings for our target devices. Next, we evaluated U-PoT's ability to mimic physical device with the following two scenarios.

\textit{\textbf{Case 1: Emulate a Known Physical Device: }}
We configured U-PoT to create an emulated version of Belkin WeMo switch that we have used for the development of the U-PoT framework. The instance of the Belkin WeMo switch was deployed in a virtual machine. The WeMo device management Android application was able to recognize the emulated IoT device. The application was unaware that the device listed by it is just an emulated device, not a real IoT physical device. We were able to operate the application to send requests to the emulated device and receive response from it. To further evaluate it, we installed OpenHAB framework with WeMo binding. OpenHAB was also able to recognize our emulate device as an WeMo switch. To strengthen the evaluation, we deployed our emulated WeMo device on a Linode server running Ubuntu-16.04 and searched the server IP address in IoT search engine Shodan\cite{shodan@io} and it was listed as an UPnP device. This analysis shows that, the U-PoT framework can successfully create an emulated IoT device from the device description of an UPnP-based IoT device and the emulated device can fool a real vendor provided control application as well as popular IoT search engines used by research community.

\textit{\textbf{Case 2: Emulate an Unknown Physical Device: }}
The target of the U-PoT framework is to emulate any UPnP device from its device description. To evaluate that, we selected Samsung Smart TV as our second device of choice. This device was not used for the implementation of the UPoT framework and the framework was totally unaware of the service descriptors of the device. Using the U-PoT framework, we were able to generate the GUPnP API based code for emulated device. We compiled the generated code against GUPnP library and deployed it in a virtual machine. Next, we installed Samsung Smart TV binding in OpenHAB framework. OpenHAB framework was able to recognize our emulated device as a Samsung Smart TV thus proving U-PoT's ability in creating emulated device from an unknown UPnP device's description.

\subsection{Response Time Analysis}
In this sub-section we perform a comparative analysis of our emulated device with respect to a real device. This analysis is useful to find out how closely U-PoT emulated device can mimic a real physical device. We leave session time and data capture performance as a future research.

\textit{\textbf{Experimental Setup: }}We configured U-PoT to create multiple emulated version of Belkin WeMo switch and deployed instances of the emulated switches in our test environment as shown in Figure \ref{fig:fig_deployment}. The test environment was created with a VirtualBox virtual machine with 1 CPU and 2GB RAM running on a Windows 10 host machine. The virtual machine was used to sandbox U-PoT emulated device environment from the host physical environment. The virtual machine was running Ubuntu-16.04 and configured with the bridge networking setting. In an actual physical environment, each device is assigned a different IP address. We created a shell script to automatically create multiple virtual interfaces on the virtual machine. Each virtual interface receives an IP address from a DHCP server located on the router. We bound an emulated device to each of this virtual interface's IP address. This flexible and configurable setup allowed us to deploy a large collection of U-PoT device within a single virtual environment with minimal effort. %In following sub-sections, we showed different test scenarios and demonstrated their performance.

\textit{\textbf{Analysis Result: }}
Figure \ref{fig:fig_response-a} and \ref{fig:fig_response-b} show a comparison of the response time for both physical and U-PoT emulated IoT device when 10 UPnP actions are invoked from an Android device and computer, respectively. As shown in the figure, the response times for emulated devices are faster than the physical devices. This is because of the fact that the physical device has limited processing capability in terms of hardware used. On the other hand, U-PoT emulated IoT device is running on a laptop with faster processing power. Though it seems that an attacker can use the response time to differentiate a honeypot from a real system, in a real deployment scenario, the response time depends on different factors such as network bandwidth, client's processing power etc.; and, hence, merely using response time as a differentiating factor is not enough. Furthermore, a response time can be superficially inflated depending on the deployment settings.

\noindent \textit{\textbf{Scalability Analysis: }}
To demonstrate the scalability of the proposed system, we gradually increased the deployed number of U-PoT instances and sent concurrent traffic to all of them (Fig:\ref{fig:fig_scalibility-a} \& \ref{fig:fig_scalibility-b}) or a subset of them (Fig:\ref{fig:fig_scalibility-c} \& \ref{fig:fig_scalibility-d}). As Figures \ref{fig:fig_scalibility-a} \& \ref{fig:fig_scalibility-b} show, the maximum value of response time increases within a reasonable limit with increased number of U-PoT honeypots if there are concurrent requests to all the honeypots. On the other hand, Figures \ref{fig:fig_scalibility-c} and \ref{fig:fig_scalibility-d} show that for concurrent requests to a constant number of honeypots (5 in our experiments); the maximum response time remains almost constant. The response time is only affected by the U-PoT devices that are serving client requests.
%As the idle U-PoT devices take minimal processing power, they do not put much overhead on the response time of the active U-PoT devices.

\section{Conclusion}
In this work, we have introduced a novel open-source honeypot framework for UPnP-based IoT devices called U-PoT. U-PoT aims to solve the problems of existing low-interaction honeypots while providing the advantages of high-interaction honeypots. It is agnostic of device type or vendor, flexible, and easily configurable for any UPnP-based devices. Specifically, we developed and introduced the architecture of U-PoT as a deploy-ready honeypot for UPnP-based smart home devices. In U-PoT, we also developed a mechanism to automatically create the honeypot from device description documents of an UPnP device (i.e., WeMo smart switch). Our evaluation shows that U-PoT can emulate a real physical IoT device successfully and can fool a real device management/control application (i.e., actual vendor-supplied Smartphone App) provided by the device vendor. Also, multiple instances of U-PoT IoT devices can be deployed in a single computer with minimum overhead on the response time. To the best of our knowledge, this is the first work to focus on the interactive nature of IoT devices to provide a cheap, flexible, and configurable honeypot framework for UPnP-based IoT devices. As a future work, we intend to deploy U-PoT in a larger public network and evaluate its average session time and data capture performance.

%\clearpage

% or


%\appendices
%\blindtext

\section*{Acknowledgment}
This work was partially supported by US NSF-CAREER-CNS-1453647 and Florida Center for Cybersecurity’s Capacity Building Program. The views expressed are
those of the authors only.


\ifCLASSOPTIONcaptionsoff
  \newpage
\fi



%\bibliographystyle{IEEEtran}

%\clearpage
\bibliographystyle{plain}
\bibliography{bib_papers} 

\begin{thebibliography}{10}

\bibitem{akiyama2010design}
Mitsuaki Akiyama, Makoto Iwamura, Yuhei Kawakoya, Kazufumi Aoki, and Mitsutaka
  Itoh.
\newblock Design and implementation of high interaction client honeypot for
  drive-by-download attacks.
\newblock {\em IEICE transactions on communications}, 93(5):1131--1139, 2010.

\bibitem{alosefer2010honeyware}
Yaser Alosefer and Omer Rana.
\newblock Honeyware: a web-based low interaction client honeypot.
\newblock In {\em Software Testing, Verification, and Validation Workshops
  (ICSTW), 2010 Third International Conference on}, pages 410--417. IEEE, 2010.

\bibitem{antonakakis2017understanding}
Manos Antonakakis, Tim April, Michael Bailey, Matt Bernhard, Elie Bursztein,
  Jaime Cochran, Zakir Durumeric, J~Alex Halderman, Luca Invernizzi, Michalis
  Kallitsis, et~al.
\newblock Understanding the mirai botnet.
\newblock In {\em USENIX Security Symposium}, 2017.

\bibitem{okiru}
Ionut Arghire.
\newblock Mirai variant targets arc cpu-based devices.
\newblock
  https://www.infosecurity-magazine.com/news/massive-qbot-strikes-500000-pcs/,
  Jan 2016.

\bibitem{baecher2006nepenthes}
Paul Baecher, Markus Koetter, Thorsten Holz, Maximillian Dornseif, and Felix
  Freiling.
\newblock The nepenthes platform: An efficient approach to collect malware.
\newblock In {\em International Workshop on Recent Advances in Intrusion
  Detection}, pages 165--184. Springer, 2006.

\bibitem{brzeczko2014active}
Albert Brzeczko, A~Selcuk Uluagac, Raheem Beyah, and John Copeland.
\newblock Active deception model for securing cloud infrastructure.
\newblock In {\em Computer Communications Workshops (INFOCOM WKSHPS), 2014 IEEE
  Conference on}, pages 535--540. IEEE, 2014.

\bibitem{durumeric2013zmap}
Zakir Durumeric, Eric Wustrow, and J~Alex Halderman.
\newblock Zmap: Fast internet-wide scanning and its security applications.
\newblock In {\em USENIX Security Symposium}, volume~8, pages 47--53, 2013.

\bibitem{upnp_architecture}
UPnP Forum.
\newblock {UPnP} upnp device architecture.
\newblock http://www.upnp.org/specs/arch/UPnP-arch-DeviceArchitecture-v1.1.pdf.

\bibitem{gupnp}
Gnome.
\newblock {}.
\newblock https://github.com/GNOME/gupnp.

\bibitem{guarnizo2017siphon}
Juan~David Guarnizo, Amit Tambe, Suman~Sankar Bhunia, Mart{\'\i}n Ochoa,
  Nils~Ole Tippenhauer, Asaf Shabtai, and Yuval Elovici.
\newblock Siphon: Towards scalable high-interaction physical honeypots.
\newblock In {\em Proceedings of the 3rd ACM Workshop on Cyber-Physical System
  Security}, pages 57--68. ACM, 2017.

\bibitem{belkin}
Belkin International.
\newblock {} belkin wemo.
\newblock https://www.belkin.com/us/.

\bibitem{luo2017iotcandyjar}
Tongbo Luo, Zhaoyan Xu, Xing Jin, Yanhui Jia, and Xin Ouyang.
\newblock Iotcandyjar: Towards an intelligent-interaction honeypot for iot
  devices.
\newblock {\em Black Hat}, 2017.

\bibitem{moore2013security}
H~Moore.
\newblock Security flaws in universal plug and play: Unplug. don’t play.
\newblock {\em Rapid7, Ltd}, 8, 2013.

\bibitem{muncaster_2014}
Phil Muncaster.
\newblock Massive qbot botnet strikes 500,000 machines through wordpress.
\newblock
  https://www.infosecurity-magazine.com/news/massive-qbot-strikes-500000-pcs/.

\bibitem{nazario2009phoneyc}
Jose Nazario.
\newblock Phoneyc: A virtual client honeypot.
\newblock {\em LEET}, 9:911--919, 2009.

\bibitem{openHAB}
openHAB.
\newblock {} open home automation bus.
\newblock https://www.openhab.org/.

\bibitem{pa2015iotpot}
Yin Minn~Pa Pa, Shogo Suzuki, Katsunari Yoshioka, Tsutomu Matsumoto, Takahiro
  Kasama, and Christian Rossow.
\newblock Iotpot: analysing the rise of iot compromises.
\newblock {\em EMU}, 9:1, 2015.

\bibitem{pauna2014casshh}
Adrian Pauna and Victor~Valeriu Patriciu.
\newblock Casshh--case adaptive ssh honeypot.
\newblock In {\em International Conference on Security in Computer Networks and
  Distributed Systems}, pages 322--333. Springer, 2014.

\bibitem{provos2004virtual}
Niels Provos et~al.
\newblock A virtual honeypot framework.
\newblock In {\em USENIX Security Symposium}, volume 173, pages 1--14, 2004.

\bibitem{seifert2007honeyc}
Christian Seifert, Ian Welch, Peter Komisarczuk, et~al.
\newblock Honeyc-the low-interaction client honeypot.
\newblock {\em Proceedings of the 2007 NZCSRCS, Waikato University, Hamilton,
  New Zealand}, 6, 2007.

\bibitem{vsemic2017iot}
Haris {\v{S}}emi{\'c} and Sasa Mrdovic.
\newblock Iot honeypot: A multi-component solution for handling manual and
  mirai-based attacks.
\newblock In {\em Telecommunication Forum (TELFOR), 2017 25th}, pages 1--4.
  IEEE, 2017.

\bibitem{shodan@io}
Shodan.io.
\newblock {} iot search engine.
\newblock https://www.shodan.io/.

\bibitem{wagener2011adaptive}
G{\'e}rard Wagener, Radu State, Thomas Engel, and Alexandre Dulaunoy.
\newblock Adaptive and self-configurable honeypots.
\newblock In {\em Integrated Network Management (IM), 2011 IFIP/IEEE
  International Symposium on}, pages 345--352. IEEE, 2011.

\bibitem{wang2017thingpot}
Meng Wang, Javier Santillan, and Fernando Kuipers.
\newblock Thingpot: an interactive internet-of-things honeypot.
\newblock 2017.

\end{thebibliography}

%\begin{biography}[{\includegraphics[width=1in,height=1.25in,clip,keepaspectratio]{mshell}}]{Michael Shell}

\end{document}